\documentclass[%
 reprint,
 superscriptaddress,
 nobibnotes,
 amsmath,amssymb,
 prl,
 twocolumn,
 dblfloatfix,
]{revtex4-1}

\usepackage[usenames,dvipsnames,svgnames,table]{xcolor}
\colorlet{m}{black}
\colorlet{k}{Black}

\usepackage{graphicx}
\usepackage[outdir=./]{epstopdf}

\usepackage{dcolumn}
\usepackage{bm}
\usepackage[colorlinks=true,citecolor=blue]{hyperref}
\usepackage[rightcaption]{sidecap}

\begin{document}

\title{Defeating broken symmetry with doping: Symmetric resonant tunneling\\ in noncentrosymetric heterostructures}
\author{Jimy Encomendero}
\email{jje64@cornell.edu}
\affiliation{School of Electrical and Computer Engineering, Cornell University, Ithaca NY 14853 USA}%

\author{Vladimir Protasenko}
\affiliation{School of Electrical and Computer Engineering, Cornell University, Ithaca NY 14853 USA}%

\author{Debdeep Jena}
\email{djena@cornell.edu}
\affiliation{School of Electrical and Computer Engineering, Cornell University, Ithaca NY 14853 USA}%
\affiliation{Department of Materials Science and Engineering, Cornell University, Ithaca NY 14853 USA}%

\author{Huili Grace Xing}
\email{grace.xing@cornell.edu}
\affiliation{School of Electrical and Computer Engineering, Cornell University, Ithaca NY 14853 USA}%
\affiliation{Department of Materials Science and Engineering, Cornell University, Ithaca NY 14853 USA}%
\affiliation{\hbox{Kavli Institute at Cornell for Nanoscale Science, Cornell University, Ithaca, New York 14853, USA}}%


\begin{abstract}
Resonant tunneling transport in polar heterostructures is intimately connected to the polarization fields emerging from the geometric Berry-phase. In these structures, quantum confinement results not only in a discrete electronic spectrum, but also in built-in polarization charges exhibiting a broken inversion symmetry along the transport direction. Thus, electrons undergo highly asymmetric quantum interference effects with respect to the direction of current flow. By employing doping to counter the broken symmetry, we deterministically control the resonant transmission through GaN/AlN resonant tunneling diodes and experimentally demonstrate the recovery of symmetric resonant tunneling injection across the noncentrosymmetric double-barrier potential.

\end{abstract}
\maketitle

\section{I. Introduction}
Tunneling is an inherent quantum mechanical phenomenon, allowing particles to propagate through classically forbidden regions of space. The experimental observation of this effect in multiple physical phenomena provided a first glimpse at the intrinsic wave properties of matter~\cite{Oppenheimer1928,Gamow1928,Zener1934,Esaki1958}. Counterintuitively, though a single tunneling barrier exponentially attenuates the tunneling particles; a double-barrier potential can allow perfect transmission. This phenomenon emerges from constructive quantum interference of the electron waves within the double-barrier cavity, provided that the particle wavelength is on the order of the barrier spacing~\cite{Tsu1973}. 

Coherent quantum interference effects in resonant tunneling cavities have been studied over the last five decades in semiconducting~\cite{Chang1974,Tsuchiya1987,Reuscher1996,Slobodskyy2003,Patane2010}, metallic~\cite{Tao2019}, ferroelectric~\cite{Du2014,Su2021} and multiferroic~\cite{Sanchez2017} materials. These effects play a central role in shaping the electronic wavefunctions, thereby controlling the transport dynamics, dipole matrix elements and scattering rates in a plethora of technologically relevant devices; including single-photon detectors~\cite{Blakesley2005}, quantum cascade lasers~\cite{Faist1994}, resonant tunneling oscillators~\cite{Yu2021,Asada2016,Xing2019,Cho2020} and transistors~\cite{Bonnefoi1985,Lind2004,Condori2016}. 

In contrast to non-polar semiconductors such as Si and AlGaAs, tunneling in polar heterostructures is dramatically modified by the internal polarization fields emerging from the geometric Berry-phase~\cite{King1993,Resta1992}. In these heterostructures, quantum confinement results in a discrete electronic spectrum, but the built-in polarization charges formed at the interfaces, introduce a broken inversion symmetry along the transport direction. As a result, electrons undergo highly asymmetric quantum interference effects with respect to the direction of current flow. Owing to the exponential relationship between the electronic wavefunction and internal electric fields, tunneling offers a unique opportunity to study Berry-phase-driven polarization effects with the highest sensitivity.

\begin{figure}[t!]
	\centerline{\includegraphics[width=0.48\textwidth]{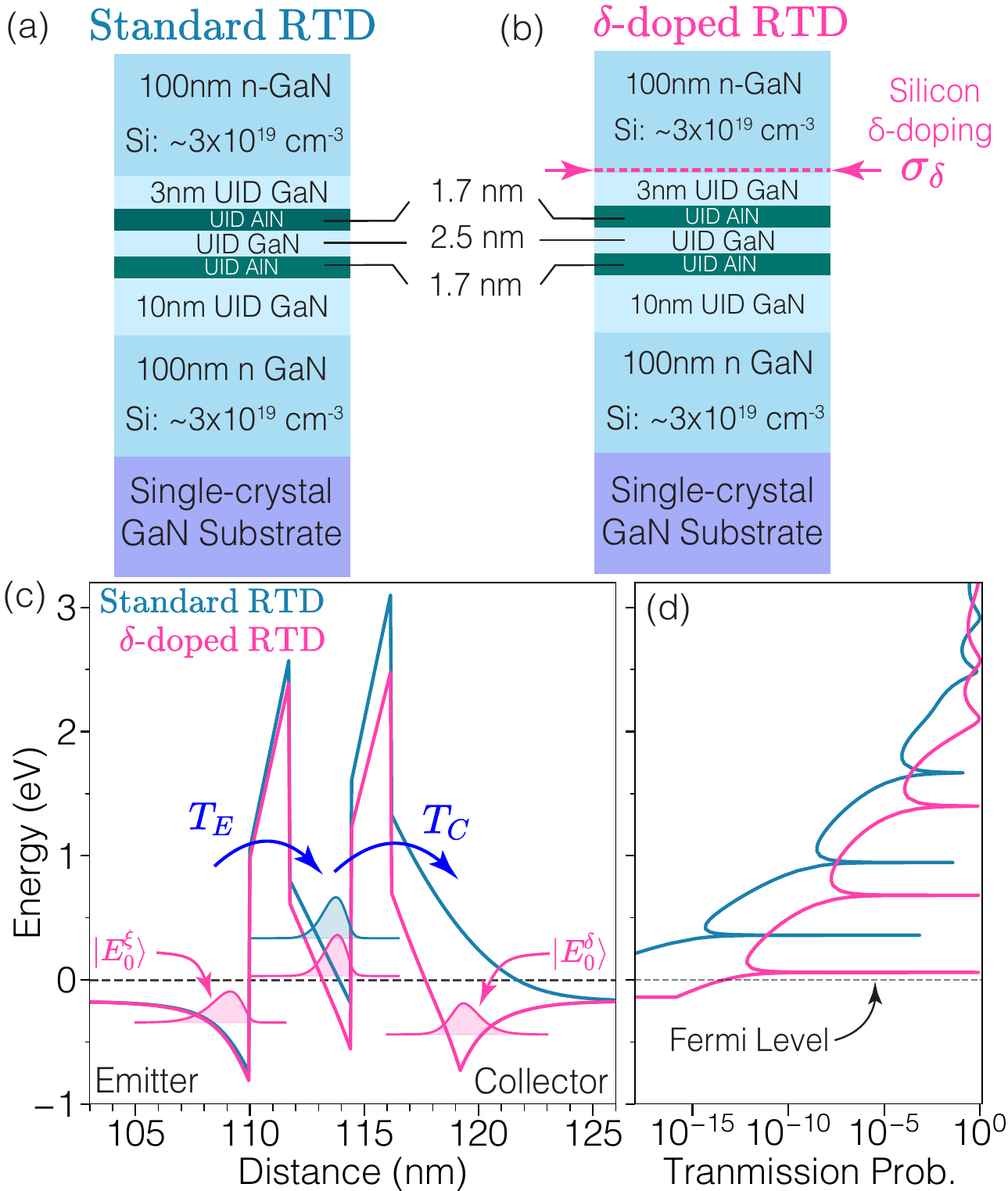}}
	\caption{Electronic quantum interference effects are studied in two different GaN/AlN double-barrier heterostructures. Both structures feature the same double-barrier active region, but different collector designs. (a) The standard RTD exhibits asymmetric band bending outside the active structure due to the broken-inversion symmetry of the polarization charges. (b) In contrast, the $\delta$-doped RTD incorporates a sheet of silicon donors $\sigma_\delta$ on the collector, resulting in an accumulation well. (c) The energy band diagram and (d) transmission probability for both structures is calculated using a self-consistent Schr\"odinger-Poisson solver and transfer matrix method.}
	\label{fig_structure}
\end{figure}

In this Letter, we employ a double-barrier resonant tunneling cavity---shown in Fig.~\ref{fig_structure}(a)---to probe polarization-induced effects on tunneling electrons. In this structure, the broken inversion symmetry of the polarization charges, gives rise not only to a strong electric polarization in the barriers and well, but also induces asymmetric band bending outside the tunneling structure [See the standard resonant tunneling diode (RTD) in Fig.~\ref{fig_structure}(c)]. As a result, electrons traversing the active region undergo highly asymmetric quantum interference effects. The broken-symmetry gives rise to nonreciprocal electronic transport with exponentially asymmetric forward ($J_p^{For}$) and reverse ($J_p^{Rev}$)  resonant tunneling peaks: $J_p^{For}/J_p^{Rev}\approx 10^7$~\cite{Encomendero2019}. Here, we experimentally demonstrate a new GaN/AlN RTD in which symmetric resonant tunneling injection is restored by countering and canceling the polarization asymmetry by doping.

\section{II. Restoring symmetric quantum interference}
Because electronic transmission is exponentially sensitive to the thickness and electric polarization of the tunneling barriers, the standard RTD structure, shown in Fig~\ref{fig_structure}(a) and (c), exhibits highly asymmetric single-barrier transmission coefficients with $T_E\gg T_C$, where $T_E$ ($T_C$) is the emitter (collector) single-barrier transmission. As a consequence, the electronic wavefunction undergoes partially constructive quantum interference, resulting in a strongly attenuated resonance transmission:  $T_\text{RES}=\frac{4T_ET_C}{(T_C+T_E)^2}\sim\frac{T_C}{T_E}\ll 1$~\cite{Ricco1984,Buttiker1988,Encomendero2019}. This effect is evident from the transmission probability displayed in Fig~\ref{fig_structure}(d), which also reveals a considerable energy shift due to quantum-confined Stark effect (QCSE) inside the GaN quantum well. 

To maximize constructive quantum interference, it is critical to restore the symmetry between the single-barrier transmission coefficients ($T_E \approx T_C$). One way to achieve this is to increase the doping concentration, thereby reducing the tunneling distance between the well and collector electrode. In this Letter, we aim to completely remove the depletion region on the collector side by inserting a $\delta$-doped layer~\cite{Schubert1994_1}, thereby exponentially enhancing the well-collector coupling strength. This new $\delta$-doped RTD concept is illustrated schematically in Fig.~\ref{fig_structure}(b). The active region is identical to the standard RTD shown in Fig.~\ref{fig_structure}(a); the only difference between the two stuctures is the incorporation of a sheet of silicon donors with density $\sigma_\delta\approx 5\times 10^{13}$~cm$^{-2}$. Comparison of the self-consistent energy-band diagrams in Fig.~\ref{fig_structure}(c) reveals that the depletion region is reduced by $\approx 7$~nm, leading to a smaller tunneling distance and exponentially enhanced wavefunction transmission. A concomitant reduction in the electric field across the AlN barriers is also evident in the $\delta$-doped RTD. In contrast, the field in the well increases, giving rise to a stronger QCSE. As a result, the ground-state energy approaches the Fermi level, thereby lowering the resonant tunneling voltage. Thanks to the improved symmetry between the tunneling barriers, electrons are expected to undergo enhanced resonant injection with: $T_\text{RES}=\frac{4T_ET_C}{(T_C+T_E)^2}\sim\frac{T_C}{T_E}\approx1$ [See Fig.~\ref{fig_structure}(d)]. Thus, owing to the enhanced tunneling transmission and low-bias resonant injection, symmetric current-voltage (\textit{J-V}) peaks and negative differential conductance (NDC) are expected under both bias polarities in the $\delta$-doped heterostructure.

\section{III. Experiment}
\begin{figure}[t]
	\includegraphics[width=0.48\textwidth]{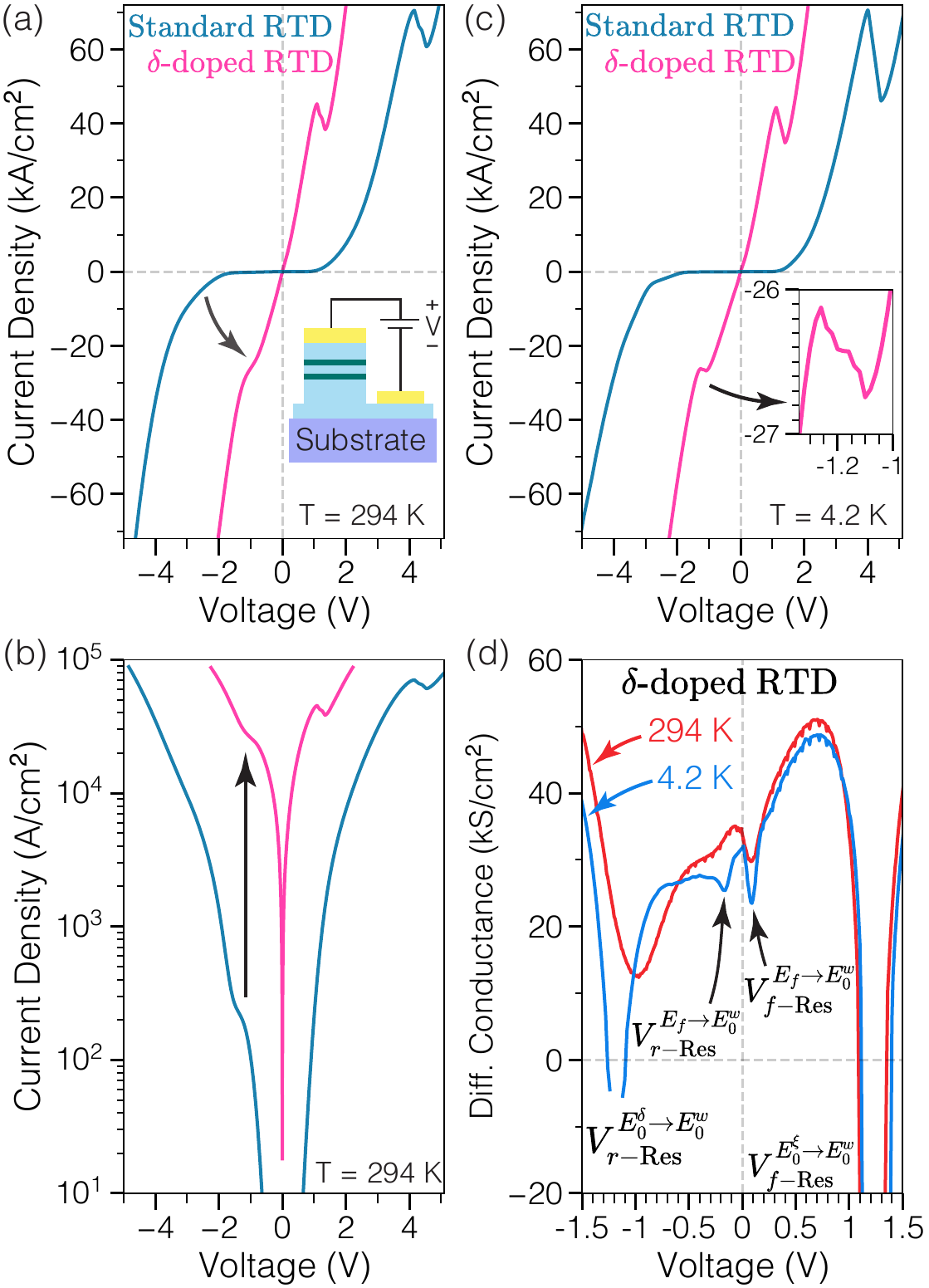}
	\caption{Electronic transport is measured through both RTD structures depicted in Fig.~\ref{fig_structure}. (a) Below resonance, the current-voltage characteristics of the $\delta$-doped design exhibits an improved linearity and symmetry with respect to the bias polarity. (b) Owing to the stronger coupling between the resonant states and the electrodes, exponentially enhanced tunneling currents are obtained under both bias polarities. (c) Tunneling transport is also measured at 4.2~K, revealing symmetric resonant tunneling injection and NDC in the $\delta$-doped RTDs, with symmetric peak voltages measured at $V_{B}=\pm 1.1$~V. (d) The temperature-dependent differential conductance of the $\delta$-doped structure reveals two conductance minima under forward and reverse current injection bias.}
	\label{fig_IVs}
\end{figure}

Molecular beam epitaxy was used to grow the heterostructures shown in Fig.~\ref{fig_structure}. Both structures were fabricated into diodes following a procedure described elsewhere~\cite{Encomendero2018,Encomendero2020,Encomendero2021_JVSTA}. The forward-biased circuit of a device under test is shown in the inset of Fig.~\ref{fig_IVs}(a). The \textit{J-V} characteristics in the same figure show that constructive quantum interference and room temperature NDC are attained in both the standard and $\delta$-doped heterostructures under forward current injection. Owing to the stronger coupling between the resonant states and electrodes, the tunneling current through the $\delta$-doped RTD is exponentially enhanced under both bias polarities [See Fig.~\ref{fig_IVs}(b)]. The cryogenic \textit{J-V} curves---measured at 4.2~K---reveal that when the $\delta$-doped structure is biased at $V_{\text{B}}=\pm1.1$~V, symmetric resonant tunneling injection and NDC is attained [Fig.~\ref{fig_IVs}(c)].

\begin{figure*}
	\includegraphics[width=\textwidth]{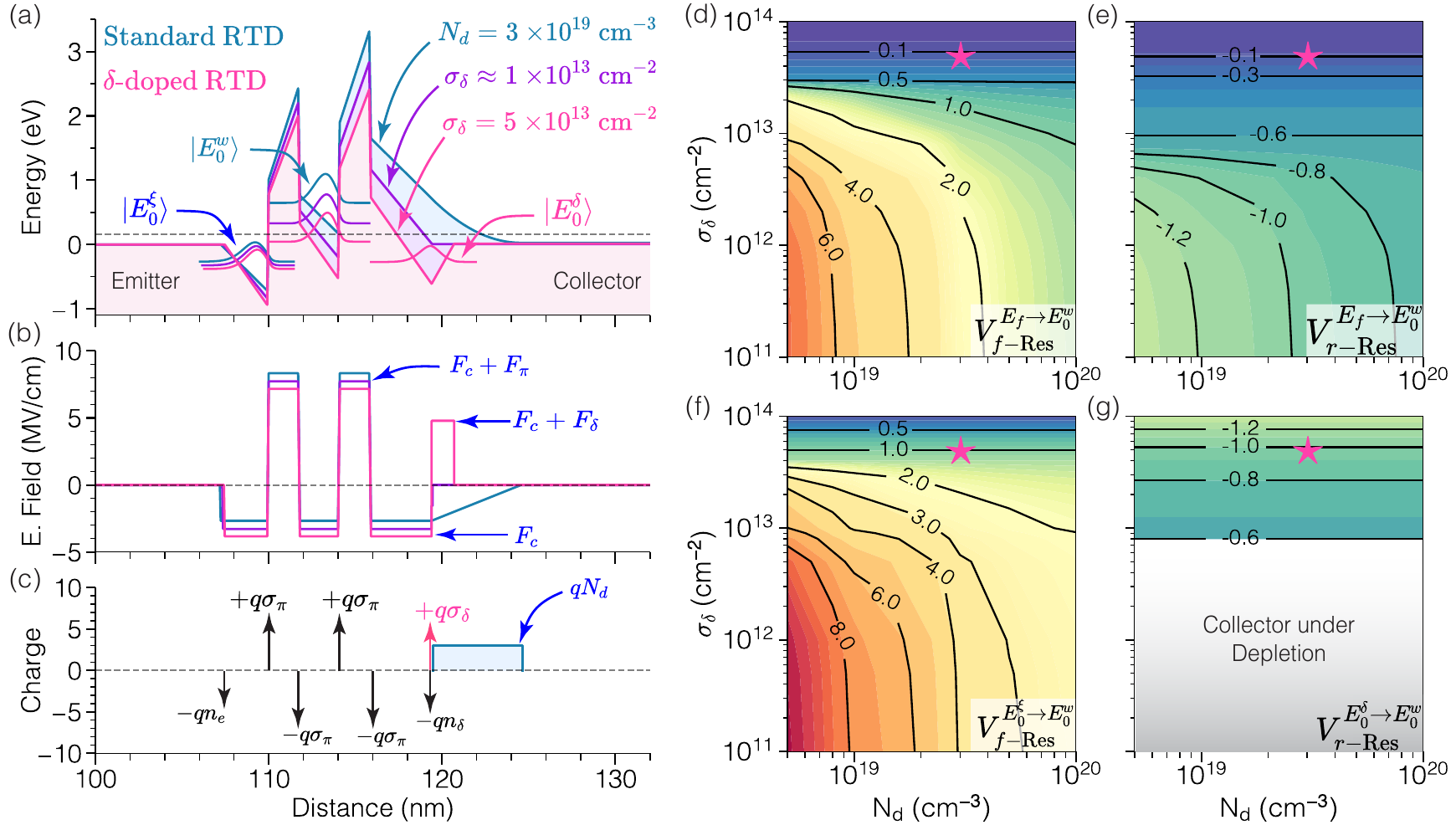}
	\caption{Self-consistent analytical model for polar double-barrier heterostructures. (a) The equilibrium band diagram, (b) electric field, and (c) charge profile are calculated for both device structures depicted in Fig.~\ref{fig_structure}. Under forward (reverse) current injection, the resonant ground-state $|E_0^w\rangle$ couples to the emitter (collector) states, enabling resonant tunneling injection across the double-barrier active region. Employing our model we calculate the voltages at which resonant coupling occurs as a function of the contact doping $N_d$ and collector delta-doping $\sigma_\delta$. Panels (d)-(g) display the theoretical (contour lines) and experimental (star) values of the resonant tunneling voltages $V^{E_f\rightarrow E_0^w}_{f-\text{Res}}$, $V^{E_f\rightarrow E^w_0}_{r-\text{Res}}$, $V^{E_0^\xi\rightarrow E_0^w}_{f-\text{Res}}$ and $V^{E_0^\delta\rightarrow E^w_0}_{r-\text{Res}}$, under both bias polarities.}
	\label{fig_model}
\end{figure*}

Moreover, in contrast to the highly nonlinear $\textit{J-V}$ curves in the standard heterostructure, transport in the $\delta$-doped RTD exhibits an improved linearity and symmetry with respect to the bias polarity. As evident from the inset of Fig.~\ref{fig_IVs}(c), the forward and reverse peak current densities are measured at $J_p\approx 44$~kA/cm$^2$, and $J_p\approx -26.7$~kA/cm$^2$, respectively. These resonances correspond to tunneling injection from the emitter and collector subbands [$|E^\xi_0\rangle$ and $|E^\delta_0\rangle$ in Fig.~\ref{fig_structure}(c)], into the resonant level, $|E_0^w\rangle$. In contrast, resonant injection from the Fermi sea occurs at a lower voltage, close to equilibrium. This can be seen from the differential conductance of the $\delta$-doped structure, shown in Fig.~\ref{fig_IVs}(d). The conductance curves, measured at different temperatures, show clear forward and reverse valleys near zero bias, indicated by the labels $V^{E_f\rightarrow E_0^w}_{f-\text{Res}}$ and $V^{E_f\rightarrow E^w_0}_{r-\text{Res}}$, respectively. 

\section{IV. $\delta$-doped RTD model}
The physical origin behind the recovery of symmetric resonant injection is explained using a self-consistent analytical model. This theoretical framework elucidates the role of the internal polarization charges in the asymmetric tunneling transport. It also sheds light into the tradeoffs within the design space of noncentrosymmetric  polar RTDs.

The collector electric field $F_c$ and the voltage bias $V_B$ for the case of the standard double-barrier heterostructure are related by the expression:
\begin{equation}
	\begin{aligned}
		V_{\text{B}}=\frac{1}{2}\frac{\varepsilon_SF_c^2}{qN_d}-F_c\left(\frac{\varepsilon_S}{C^\xi_q}+t_\text{DB}+t_s\right)-2F_\pi t_b+\frac{E_0^\xi}{q}.
	\end{aligned}
	\label{eq_std}
\end{equation}
Here, $N_d$ is the contact doping concentration, $\varepsilon_S$ is the dielectric constant of GaN, and $F_\pi=q\sigma_\pi/\varepsilon_S$ is the internal polarization field. $t_{\text{DB}}=t_b+t_w+t_b$ is the total thickness of the double-barrier active region. $t_b$, $t_w$, and $t_s$ are the barrier, well, and collector spacer thickness, respectively. The triangular accumulation well on the emitter side, hosts a bound-state $|E_0^\xi\rangle$ with energy $E_0^\xi$, measured from the bottom of the well [See Fig.~\ref{fig_model}(a)]. Employing a variational approach, we find that $E_0^\xi=\frac{3}{2}\left(\frac{3q\hbar |F_c|}{2\sqrt{m^\star}}\right)^{\frac{2}{3}}$, where $\hbar$ is the reduced Planck's constant and $m^\star$ is the conduction band-edge effective mass of GaN~\cite{Schubert1994_1}. The charge accumulation in the emitter well is captured by its quantum capacitance: $C^\xi_q=\frac{q^2 m^\star}{\pi\hbar^2}$. 

We note that the first term in Eq.~\ref{eq_std} corresponds to the voltage dropped across the depletion region, and scales quadratically with the collector field, $F_c$. Thus, the space-charge modulation within this region is responsible for the weak voltage-bias tuning of the cavity into resonance; resulting in a larger forward resonant voltage. In contrast to Eq.~\ref{eq_std}, the functional dependence of $V_B$ in the $\delta$-doped RTD scales linearly with $F_c$, and is given by: 
\begin{equation}
	\begin{aligned}
		V_{\text{B}}=-F_c\left(\frac{\varepsilon_S}{C_q^\xi}+t_\text{DB}+t_s+\frac{\varepsilon_S}{C_q^\delta}\right)-2F_\pi t_b+\\ \frac{E_0^\xi}{q}-\frac{E_0^\delta}{q}-\frac{\varepsilon_S}{C_q^\delta}F_\delta.
	\end{aligned}
	\label{eq_delta}
\end{equation}
$C^\delta_q=\frac{q^2 m^\star}{\pi\hbar^2}$ is the quantum capacitance of the collector $\delta$-well, whose ground-state eigenenergy reads $E^\delta_0=\frac{3}{10}\left(\frac{9q\hbar F_\delta}{4\sqrt{m^\star}}\right)^{\frac{2}{3}}$~\cite{Schubert1994_1}. Equations~\ref{eq_std} and~\ref{eq_delta} are of central importance in our analytical model, enabling the calculation of the energy-band diagram, electric field profile, and charge distribution for a general $\delta$-doped RTD, under equilibrium and non-equilibrium conditions [See Figs.~\ref{fig_model}(a)-(c)]. Here, we apply this unified theoretical framework to the case of non-centrosymmetric GaN/AlN RTDs.

Figure~\ref{fig_model}(a) displays the equilibrium energy-band diagrams of different GaN/AlN RTD designs, obtained using our model. The active region of these structures is identical to the devices showcased in Fig.~\ref{fig_structure}(a). Under current injection, the main resonant tunneling peaks originate from the coupling between $|E^w_0\rangle$ and the emitter (collector) bound states $|E^\xi_0\rangle$ ($|E^\delta_0\rangle$)~\cite{Encomendero2019,Encomendero2017,Encomendero2021APEX,Encomendero2022APS,Growden2020}.  Detuning from these resonant configurations gives rise to NDC valleys, labeled by $V^{E_0^\xi\rightarrow E_0^w}_{f-\text{Res}}$ and $V^{E_0^\delta\rightarrow E^w_0}_{r-\text{Res}}$ in Fig.~\ref{fig_IVs}(d).

Panels (d)-(g) in Fig.~\ref{fig_model} display the contour lines of the resonant voltages $V^{E_f\rightarrow E_0^w}_{f-\text{Res}}$, $V^{E_f\rightarrow E^w_0}_{r-\text{Res}}$, $V^{E_0^\xi\rightarrow E_0^w}_{f-\text{Res}}$ and $V^{E_0^\delta\rightarrow E^w_0}_{r-\text{Res}}$ for the structures shown in Fig.~\ref{fig_model}(a). The analytical form of our model allows the exploration of the RTD design space as a function of $N_d$ and $\sigma_\delta$, over a wide parameter range spanning several orders of magnitude. These results reveal a reasonable agreement between experimental values seen in Fig.~\ref{fig_IVs} and theoretical resonant voltages for all resonances, and under both bias polarities [See Fig.~\ref{fig_model}(d)-(g)]. As evident from Figs.~\ref{fig_model}(d) and (f), a monotonic reduction in the forward resonant voltages is obtained as $\sigma_\delta$ increases over several orders of magnitude. Moreover, for doping levels $\sigma_\delta \lessapprox 3\times 10^{13}$~cm$^{-2}$, the forward resonant voltages exhibit a strong dependence on the doping concentration $N_d$. This behavior stems from the inverse dependence between $N_d$ and the collector depletion required to screen the polarization charges within the double-barrier active region in Eq.~\ref{eq_std}. When the donor sheet density increases beyond $\sigma_\delta \gtrapprox 3\times 10^{13}$~cm$^{-2}$, both $V^{E_f\rightarrow E_0^w}_{f-\text{Res}}$ and $V^{E_\xi\rightarrow E_0^w}_{f-\text{Res}}$ become effectively independent of the 3D-doping level $N_d$, and are therefore independent of the internal polarization fields. Within this design subspace, only the 2D-doping level $\sigma_\delta$ controls the forward resonances [See Figs.~\ref{fig_model}(d) and (f)].

\begin{figure*}
	\includegraphics[width=0.81\textwidth]{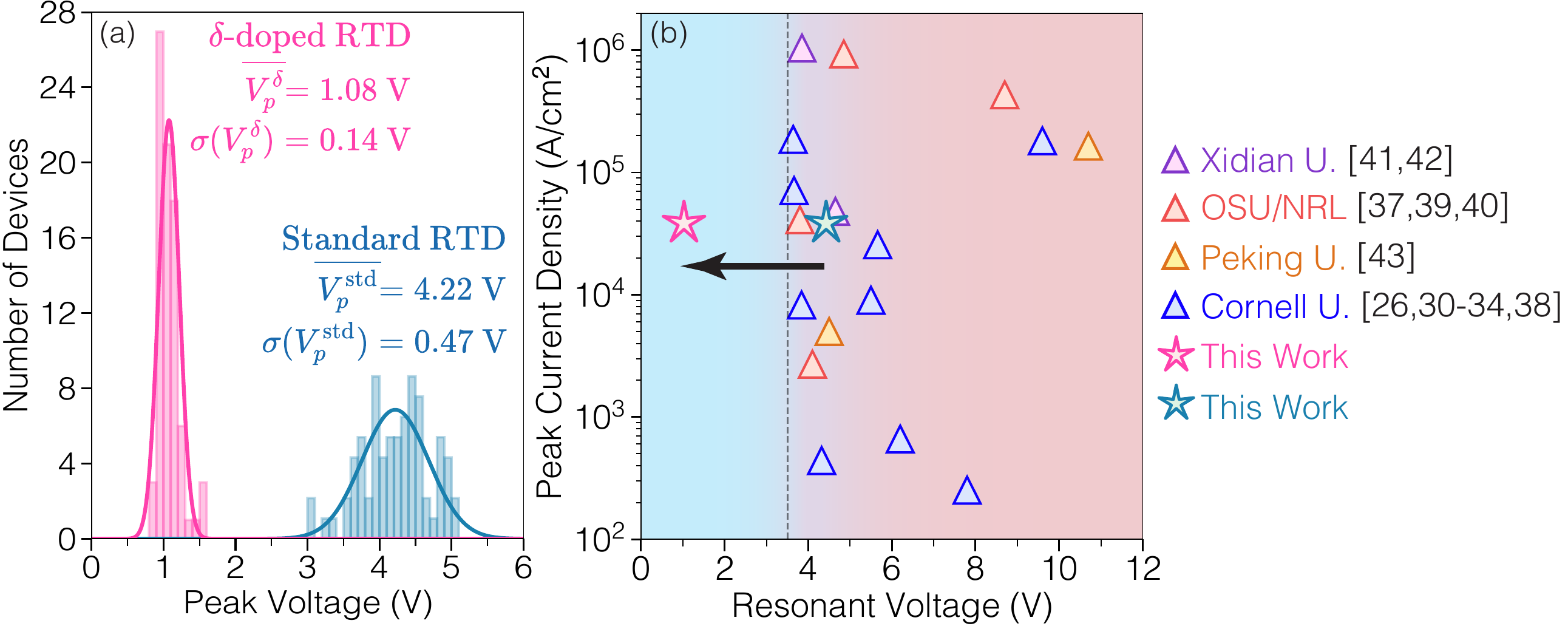}
	\caption{(a) Histogram of the peak voltage distribution obtained after measuring 81 different RTDs from each heterostructure design shown in Fig.~\ref{fig_structure}. The mean $\overline{V_p}$ and  standard deviation $\sigma\left(V_p\right)$ are displayed next to each distribution. (b) To put these results in context, we compare the resonant tunneling voltages and peak current densities presented here, with previous experimental reports of GaN/AlN RTDs~\cite{,Encomendero2018,Encomendero2020,Encomendero2021_JVSTA,Growden2016,Encomendero2019,Encomendero2021APEX,Encomendero2020PRA,Encomendero2017,Growden2020,Encomendero2022APS,Growden2018,Growden2019,Zhang2021JAP,Zhang2021APL,Wang2018}. The $\delta$-doped design, presented in this Letter, enables access to the lowest resonant tunneling voltages for the first time.}
\label{fig_bench}
\end{figure*}

In contrast to the forward injection regime, the reverse-bias resonances ($V^{E_f\rightarrow E_0^w}_{r-\text{Res}}$ and $V^{E_\delta\rightarrow E_0^w}_{r-\text{Res}}$) exhibit dramatically different trends as $\sigma_\delta$ increases [See Figs.~\ref{fig_model}(e) and (g)]. Owing to the asymmetric band bending in the standard RTD, electrons injected from the collector side are supplied only by the Fermi sea, resulting in a unique reverse resonant voltage $V^{E_f\rightarrow E_0^w}_{r-\text{Res}}$. However as $\sigma_\delta$ approaches the doping density of $1\times 10^{13}$~cm$^{-2}$, the collector depletion region is fully suppressed. This can be seen from the energy band diagram labeled as $\sigma_\delta\approx1\times 10^{13}$~cm$^{-2}$ in Fig.~\ref{fig_model}(a). Therefore, RTDs featuring this design attain resonant injection with the whole voltage bias applied across the active region. This distinctive feature enables access to the resonant configurations with the lowest voltage biases. Moreover, the suppression of the depletion region marks a transition in the design space from heterostructures with purely 3D-electron supply to a regime in which both 2D and 3D electrons are injected into the cavity. Beyond this point (i.e. $\sigma_\delta\gtrapprox 1\times 10^{13}$~cm$^{-2}$), increasing $\sigma_\delta$ results in the formation of a V-shaped $\delta$-well in the collector electrode. An example of such structure is the $\delta$-doped RTD shown in Fig.~\ref{fig_model}(a). In these RTDs, the main reverse-bias resonant peak stems from the coupling between $|E_0^w\rangle$ and $|E_0^{\delta}\rangle$. Within this design subspace and upon increasing $\sigma_\delta$, the energy of the collector subband $|E_0^{\delta}\rangle$ decreases, resulting in a more negative resonant tunneling voltage in the reverse direction. This behavior is completely captured by our model as can be seen from the larger negative values of $V^{E_0^\delta\rightarrow E_0^w}_{r-\text{Res}}$ as $\sigma_\delta\rightarrow 1\times 10^{14}$~cm$^{-2}$ [See Fig.~\ref{fig_model}(g)]. This analysis reveals that collector $\delta$-doping enables control over the quantum interference properties of polar resonant tunneling cavities, to the extent of attaining resonances independent of the internal polarization fields.

\section{V. Discussion and conclusions}
To conclusively demonstrate quantum interference control by $\delta$-doping, we measure tunneling transport across multiple RTDs with different mesa areas, varying between $25$ and $125$~$\mu$m$^2$. Peak tunneling currents from both heterostructures scale linearly with their mesa areas, revealing that electronic transport is mediated by the total area instead of the periphery. Figure~\ref{fig_bench}(a) shows the peak voltage distribution for both RTD designs obtained from a set of 81 devices per sample. Both distributions exhibit a Gaussian behavior, with their dispersion attributed to monolayer fluctuations in the thickness of the AlN barriers and GaN quantum well~\cite{Song2016}. However, the narrower spread measured in $V^{\delta}_p$, demonstrates that the $\delta$-doped design enables not only enhanced tunneling transmission and stronger electrostatic control, but also improved robustness against fluctuations in the resonant tunneling voltage. 

Finally, Fig.~\ref{fig_bench}(b) allows us to put these results in context by comparing the resonant tunneling voltages and peak current densities presented here, with previous experimental reports of GaN/AlN RTDs~\cite{Encomendero2018,Encomendero2020,Encomendero2021_JVSTA,Growden2016,Encomendero2019,Encomendero2021APEX,Encomendero2020PRA,Encomendero2017,Growden2018,Growden2019,Growden2020,Encomendero2022APS,Zhang2021JAP,Zhang2021APL,Wang2018}. Owing to the strong internal polarization in the active region, so far these devices have attained resonant tunneling injection at $V_p>3.5$~V. The $\delta$-doped design, introduced in this Letter, enables access to the low-bias resonant tunneling injection regime for the first time.

In conclusion, here we report deterministic control over the electronic quantum interference properties of noncentrosymmetric resonant tunneling heterostructures. A new $\delta$-doped RTD exhibiting symmetric constructive quantum interference under both bias polarities is unveiled. Due to the stronger coupling between the electrodes and resonant levels, the design not only delivers exponentially enhanced tunneling currents, but also enables stronger electrostatic control and improved robustness against interface roughness. The combined experimental and theoretical approach of this study opens a new avenue for the design and fabrication of robust resonant tunneling effects in noncentrosymmetric polar semiconductors.

\section{VI. Acknowledgments}
This work was funded by Office of Naval Research (ONR) under the DATE MURI Program (Contract: N00014-11-10721, Program Manager: Dr. P. Maki) and the National Science Foundation (NSF), through the MRSEC program (DMR-1719875). Support from NSF through DMREF (DMR-1534303) and EFRI-NewLAW (EFMA-1741694) programs is also acknowledged. This work was performed in part at the Cornell NanoScale Facility, supported by NSF Grant NNCI-2025233, and Cornell Center for Materials Research Shared Facilities supported through the National Science Foundation (NSF) MRSEC program (DMR-1719875).

\bibliography{z_refs}

\begin{thebibliography}{44}%
\makeatletter
\providecommand \@ifxundefined [1]{%
 \@ifx{#1\undefined}
}%
\providecommand \@ifnum [1]{%
 \ifnum #1\expandafter \@firstoftwo
 \else \expandafter \@secondoftwo
 \fi
}%
\providecommand \@ifx [1]{%
 \ifx #1\expandafter \@firstoftwo
 \else \expandafter \@secondoftwo
 \fi
}%
\providecommand \natexlab [1]{#1}%
\providecommand \enquote  [1]{``#1''}%
\providecommand \bibnamefont  [1]{#1}%
\providecommand \bibfnamefont [1]{#1}%
\providecommand \citenamefont [1]{#1}%
\providecommand \href@noop [0]{\@secondoftwo}%
\providecommand \href [0]{\begingroup \@sanitize@url \@href}%
\providecommand \@href[1]{\@@startlink{#1}\@@href}%
\providecommand \@@href[1]{\endgroup#1\@@endlink}%
\providecommand \@sanitize@url [0]{\catcode `\\12\catcode `\$12\catcode
  `\&12\catcode `\#12\catcode `\^12\catcode `\_12\catcode `\%12\relax}%
\providecommand \@@startlink[1]{}%
\providecommand \@@endlink[0]{}%
\providecommand \url  [0]{\begingroup\@sanitize@url \@url }%
\providecommand \@url [1]{\endgroup\@href {#1}{\urlprefix }}%
\providecommand \urlprefix  [0]{URL }%
\providecommand \Eprint [0]{\href }%
\providecommand \doibase [0]{http://dx.doi.org/}%
\providecommand \selectlanguage [0]{\@gobble}%
\providecommand \bibinfo  [0]{\@secondoftwo}%
\providecommand \bibfield  [0]{\@secondoftwo}%
\providecommand \translation [1]{[#1]}%
\providecommand \BibitemOpen [0]{}%
\providecommand \bibitemStop [0]{}%
\providecommand \bibitemNoStop [0]{.\EOS\space}%
\providecommand \EOS [0]{\spacefactor3000\relax}%
\providecommand \BibitemShut  [1]{\csname bibitem#1\endcsname}%
\let\auto@bib@innerbib\@empty
\bibitem [{\citenamefont {Oppenheimer}(1928)}]{Oppenheimer1928}%
  \BibitemOpen
  \bibfield  {author} {\bibinfo {author} {\bibfnamefont {J.~R.}\ \bibnamefont
  {Oppenheimer}},\ }\href {\doibase 10.1103/PhysRev.31.66} {\bibfield
  {journal} {\bibinfo  {journal} {Physical Review}\ }\textbf {\bibinfo {volume}
  {31}},\ \bibinfo {pages} {66–81} (\bibinfo {year} {1928})}\BibitemShut
  {NoStop}%
\bibitem [{\citenamefont {Gamow}(1928)}]{Gamow1928}%
  \BibitemOpen
  \bibfield  {author} {\bibinfo {author} {\bibfnamefont {G.}~\bibnamefont
  {Gamow}},\ }\href@noop {} {\bibfield  {journal} {\bibinfo  {journal}
  {Nature}\ }\textbf {\bibinfo {volume} {122}},\ \bibinfo {pages} {805–806}
  (\bibinfo {year} {1928})}\BibitemShut {NoStop}%
\bibitem [{\citenamefont {Zener}(1934)}]{Zener1934}%
  \BibitemOpen
  \bibfield  {author} {\bibinfo {author} {\bibfnamefont {C.}~\bibnamefont
  {Zener}},\ }\href {\doibase 10.1098/rspa.1934.0116} {\bibfield  {journal}
  {\bibinfo  {journal} {Proceedings of the Royal Society A: Mathematical,
  Physical and Engineering Sciences}\ }\textbf {\bibinfo {volume} {145}},\
  \bibinfo {pages} {523–529} (\bibinfo {year} {1934})}\BibitemShut {NoStop}%
\bibitem [{\citenamefont {Esaki}(1958)}]{Esaki1958}%
  \BibitemOpen
  \bibfield  {author} {\bibinfo {author} {\bibfnamefont {L.}~\bibnamefont
  {Esaki}},\ }\href {\doibase 10.1103/PhysRev.109.603} {\bibfield  {journal}
  {\bibinfo  {journal} {Physical Review}\ }\textbf {\bibinfo {volume} {109}},\
  \bibinfo {pages} {603–604} (\bibinfo {year} {1958})}\BibitemShut {NoStop}%
\bibitem [{\citenamefont {Tsu}\ and\ \citenamefont {Esaki}(1973)}]{Tsu1973}%
  \BibitemOpen
  \bibfield  {author} {\bibinfo {author} {\bibfnamefont {R.}~\bibnamefont
  {Tsu}}\ and\ \bibinfo {author} {\bibfnamefont {L.}~\bibnamefont {Esaki}},\
  }\href
  {http://scitation.aip.org/content/aip/journal/apl/22/11/10.1063/1.1654509}
  {\bibfield  {journal} {\bibinfo  {journal} {Appl. Phys. Lett.}\ }\textbf
  {\bibinfo {volume} {22}},\ \bibinfo {pages} {562} (\bibinfo {year}
  {1973})}\BibitemShut {NoStop}%
\bibitem [{\citenamefont {Chang}\ \emph {et~al.}(1974)\citenamefont {Chang},
  \citenamefont {Esaki},\ and\ \citenamefont {Tsu}}]{Chang1974}%
  \BibitemOpen
  \bibfield  {author} {\bibinfo {author} {\bibfnamefont {L.~L.}\ \bibnamefont
  {Chang}}, \bibinfo {author} {\bibfnamefont {L.}~\bibnamefont {Esaki}}, \ and\
  \bibinfo {author} {\bibfnamefont {R.}~\bibnamefont {Tsu}},\ }\href {\doibase
  10.1063/1.1655067} {\bibfield  {journal} {\bibinfo  {journal} {Appl. Phys.
  Lett.}\ }\textbf {\bibinfo {volume} {24}},\ \bibinfo {pages} {593} (\bibinfo
  {year} {1974})}\BibitemShut {NoStop}%
\bibitem [{\citenamefont {Tsuchiya}\ \emph {et~al.}(1987)\citenamefont
  {Tsuchiya}, \citenamefont {Matsusue},\ and\ \citenamefont
  {Sakaki}}]{Tsuchiya1987}%
  \BibitemOpen
  \bibfield  {author} {\bibinfo {author} {\bibfnamefont {M.}~\bibnamefont
  {Tsuchiya}}, \bibinfo {author} {\bibfnamefont {T.}~\bibnamefont {Matsusue}},
  \ and\ \bibinfo {author} {\bibfnamefont {H.}~\bibnamefont {Sakaki}},\ }\href
  {\doibase 10.1103/PhysRevLett.59.2356} {\bibfield  {journal} {\bibinfo
  {journal} {Physical Review Letters}\ }\textbf {\bibinfo {volume} {59}},\
  \bibinfo {pages} {2356–2359} (\bibinfo {year} {1987})}\BibitemShut
  {NoStop}%
\bibitem [{\citenamefont {Reuscher}\ \emph {et~al.}(1996)\citenamefont
  {Reuscher}, \citenamefont {Keim}, \citenamefont {Fischer}, \citenamefont
  {Waag},\ and\ \citenamefont {Landwehr}}]{Reuscher1996}%
  \BibitemOpen
  \bibfield  {author} {\bibinfo {author} {\bibfnamefont {G.}~\bibnamefont
  {Reuscher}}, \bibinfo {author} {\bibfnamefont {M.}~\bibnamefont {Keim}},
  \bibinfo {author} {\bibfnamefont {F.}~\bibnamefont {Fischer}}, \bibinfo
  {author} {\bibfnamefont {A.}~\bibnamefont {Waag}}, \ and\ \bibinfo {author}
  {\bibfnamefont {G.}~\bibnamefont {Landwehr}},\ }\href {\doibase
  10.1103/PhysRevB.53.16414} {\bibfield  {journal} {\bibinfo  {journal} {Phys.
  Rev. B}\ }\textbf {\bibinfo {volume} {53}},\ \bibinfo {pages} {16414–16419}
  (\bibinfo {year} {1996})}\BibitemShut {NoStop}%
\bibitem [{\citenamefont {Slobodskyy}\ \emph {et~al.}(2003)\citenamefont
  {Slobodskyy}, \citenamefont {Gould}, \citenamefont {Slobodskyy},
  \citenamefont {Becker}, \citenamefont {Schmidt},\ and\ \citenamefont
  {Molenkamp}}]{Slobodskyy2003}%
  \BibitemOpen
  \bibfield  {author} {\bibinfo {author} {\bibfnamefont {A.}~\bibnamefont
  {Slobodskyy}}, \bibinfo {author} {\bibfnamefont {C.}~\bibnamefont {Gould}},
  \bibinfo {author} {\bibfnamefont {T.}~\bibnamefont {Slobodskyy}}, \bibinfo
  {author} {\bibfnamefont {C.~R.}\ \bibnamefont {Becker}}, \bibinfo {author}
  {\bibfnamefont {G.}~\bibnamefont {Schmidt}}, \ and\ \bibinfo {author}
  {\bibfnamefont {L.~W.}\ \bibnamefont {Molenkamp}},\ }\href {\doibase
  10.1103/PhysRevLett.90.246601} {\bibfield  {journal} {\bibinfo  {journal}
  {Physical Review Letters}\ }\textbf {\bibinfo {volume} {90}},\ \bibinfo
  {pages} {246601} (\bibinfo {year} {2003})}\BibitemShut {NoStop}%
\bibitem [{\citenamefont {Patanè}\ \emph {et~al.}(2010)\citenamefont
  {Patanè}, \citenamefont {Mori}, \citenamefont {Makarovsky}, \citenamefont
  {Eaves}, \citenamefont {Zambrano}, \citenamefont {Arce}, \citenamefont
  {Dickinson},\ and\ \citenamefont {Maude}}]{Patane2010}%
  \BibitemOpen
  \bibfield  {author} {\bibinfo {author} {\bibfnamefont {A.}~\bibnamefont
  {Patanè}}, \bibinfo {author} {\bibfnamefont {N.}~\bibnamefont {Mori}},
  \bibinfo {author} {\bibfnamefont {O.}~\bibnamefont {Makarovsky}}, \bibinfo
  {author} {\bibfnamefont {L.}~\bibnamefont {Eaves}}, \bibinfo {author}
  {\bibfnamefont {M.~L.}\ \bibnamefont {Zambrano}}, \bibinfo {author}
  {\bibfnamefont {J.~C.}\ \bibnamefont {Arce}}, \bibinfo {author}
  {\bibfnamefont {L.}~\bibnamefont {Dickinson}}, \ and\ \bibinfo {author}
  {\bibfnamefont {D.~K.}\ \bibnamefont {Maude}},\ }\href {\doibase
  10.1103/PhysRevLett.105.236804} {\bibfield  {journal} {\bibinfo  {journal}
  {Physical Review Letters}\ }\textbf {\bibinfo {volume} {105}},\ \bibinfo
  {pages} {236804} (\bibinfo {year} {2010})}\BibitemShut {NoStop}%
\bibitem [{\citenamefont {Tao}\ \emph {et~al.}(2019)\citenamefont {Tao},
  \citenamefont {Wan}, \citenamefont {Tang}, \citenamefont {Feng},
  \citenamefont {Wei}, \citenamefont {Wang}, \citenamefont {Andrieu},
  \citenamefont {Yang}, \citenamefont {Chshiev}, \citenamefont {Devaux},
  \citenamefont {Hauet}, \citenamefont {Montaigne}, \citenamefont {Mangin},
  \citenamefont {Hehn}, \citenamefont {Lacour}, \citenamefont {Han},\ and\
  \citenamefont {Lu}}]{Tao2019}%
  \BibitemOpen
  \bibfield  {author} {\bibinfo {author} {\bibfnamefont {B.}~\bibnamefont
  {Tao}}, \bibinfo {author} {\bibfnamefont {C.}~\bibnamefont {Wan}}, \bibinfo
  {author} {\bibfnamefont {P.}~\bibnamefont {Tang}}, \bibinfo {author}
  {\bibfnamefont {J.}~\bibnamefont {Feng}}, \bibinfo {author} {\bibfnamefont
  {H.}~\bibnamefont {Wei}}, \bibinfo {author} {\bibfnamefont {X.}~\bibnamefont
  {Wang}}, \bibinfo {author} {\bibfnamefont {S.}~\bibnamefont {Andrieu}},
  \bibinfo {author} {\bibfnamefont {H.}~\bibnamefont {Yang}}, \bibinfo {author}
  {\bibfnamefont {M.}~\bibnamefont {Chshiev}}, \bibinfo {author} {\bibfnamefont
  {X.}~\bibnamefont {Devaux}}, \bibinfo {author} {\bibfnamefont
  {T.}~\bibnamefont {Hauet}}, \bibinfo {author} {\bibfnamefont
  {F.}~\bibnamefont {Montaigne}}, \bibinfo {author} {\bibfnamefont
  {S.}~\bibnamefont {Mangin}}, \bibinfo {author} {\bibfnamefont
  {M.}~\bibnamefont {Hehn}}, \bibinfo {author} {\bibfnamefont {D.}~\bibnamefont
  {Lacour}}, \bibinfo {author} {\bibfnamefont {X.}~\bibnamefont {Han}}, \ and\
  \bibinfo {author} {\bibfnamefont {Y.}~\bibnamefont {Lu}},\ }\href {\doibase
  10.1021/acs.nanolett.9b00205} {\bibfield  {journal} {\bibinfo  {journal}
  {Nano Letters}\ }\textbf {\bibinfo {volume} {19}},\ \bibinfo {pages} {3019}
  (\bibinfo {year} {2019})}\BibitemShut {NoStop}%
\bibitem [{\citenamefont {Du}\ \emph {et~al.}(2014)\citenamefont {Du},
  \citenamefont {Qiu}, \citenamefont {Li},\ and\ \citenamefont {Wu}}]{Du2014}%
  \BibitemOpen
  \bibfield  {author} {\bibinfo {author} {\bibfnamefont {R.}~\bibnamefont
  {Du}}, \bibinfo {author} {\bibfnamefont {X.}~\bibnamefont {Qiu}}, \bibinfo
  {author} {\bibfnamefont {A.}~\bibnamefont {Li}}, \ and\ \bibinfo {author}
  {\bibfnamefont {D.}~\bibnamefont {Wu}},\ }\href {\doibase 10.1063/1.4871277}
  {\bibfield  {journal} {\bibinfo  {journal} {Applied Physics Letters}\
  }\textbf {\bibinfo {volume} {104}} (\bibinfo {year} {2014}),\
  10.1063/1.4871277}\BibitemShut {NoStop}%
\bibitem [{\citenamefont {Su}\ \emph {et~al.}(2021)\citenamefont {Su},
  \citenamefont {Zheng}, \citenamefont {Wen}, \citenamefont {Li}, \citenamefont
  {Xie}, \citenamefont {Rabe}, \citenamefont {Liu},\ and\ \citenamefont
  {Tsymbal}}]{Su2021}%
  \BibitemOpen
  \bibfield  {author} {\bibinfo {author} {\bibfnamefont {J.}~\bibnamefont
  {Su}}, \bibinfo {author} {\bibfnamefont {X.}~\bibnamefont {Zheng}}, \bibinfo
  {author} {\bibfnamefont {Z.}~\bibnamefont {Wen}}, \bibinfo {author}
  {\bibfnamefont {T.}~\bibnamefont {Li}}, \bibinfo {author} {\bibfnamefont
  {S.}~\bibnamefont {Xie}}, \bibinfo {author} {\bibfnamefont {K.~M.}\
  \bibnamefont {Rabe}}, \bibinfo {author} {\bibfnamefont {X.}~\bibnamefont
  {Liu}}, \ and\ \bibinfo {author} {\bibfnamefont {E.~Y.}\ \bibnamefont
  {Tsymbal}},\ }\href {\doibase 10.1103/PhysRevB.104.L060101} {\bibfield
  {journal} {\bibinfo  {journal} {Physical Review B}\ }\textbf {\bibinfo
  {volume} {104}},\ \bibinfo {pages} {L060101} (\bibinfo {year}
  {2021})}\BibitemShut {NoStop}%
\bibitem [{\citenamefont {Sanchez-Santolino}\ \emph {et~al.}(2017)\citenamefont
  {Sanchez-Santolino}, \citenamefont {Tornos}, \citenamefont
  {Hernandez-Martin}, \citenamefont {Beltran}, \citenamefont {Munuera},
  \citenamefont {Cabero}, \citenamefont {Perez-Muñoz}, \citenamefont {Ricote},
  \citenamefont {Mompean}, \citenamefont {Garcia-Hernandez}, \citenamefont
  {Sefrioui}, \citenamefont {Leon}, \citenamefont {Pennycook}, \citenamefont
  {Muñoz}, \citenamefont {Varela},\ and\ \citenamefont
  {Santamaria}}]{Sanchez2017}%
  \BibitemOpen
  \bibfield  {author} {\bibinfo {author} {\bibfnamefont {G.}~\bibnamefont
  {Sanchez-Santolino}}, \bibinfo {author} {\bibfnamefont {J.}~\bibnamefont
  {Tornos}}, \bibinfo {author} {\bibfnamefont {D.}~\bibnamefont
  {Hernandez-Martin}}, \bibinfo {author} {\bibfnamefont {J.~I.}\ \bibnamefont
  {Beltran}}, \bibinfo {author} {\bibfnamefont {C.}~\bibnamefont {Munuera}},
  \bibinfo {author} {\bibfnamefont {M.}~\bibnamefont {Cabero}}, \bibinfo
  {author} {\bibfnamefont {A.}~\bibnamefont {Perez-Muñoz}}, \bibinfo {author}
  {\bibfnamefont {J.}~\bibnamefont {Ricote}}, \bibinfo {author} {\bibfnamefont
  {F.}~\bibnamefont {Mompean}}, \bibinfo {author} {\bibfnamefont
  {M.}~\bibnamefont {Garcia-Hernandez}}, \bibinfo {author} {\bibfnamefont
  {Z.}~\bibnamefont {Sefrioui}}, \bibinfo {author} {\bibfnamefont
  {C.}~\bibnamefont {Leon}}, \bibinfo {author} {\bibfnamefont {S.~J.}\
  \bibnamefont {Pennycook}}, \bibinfo {author} {\bibfnamefont {M.~C.}\
  \bibnamefont {Muñoz}}, \bibinfo {author} {\bibfnamefont {M.}~\bibnamefont
  {Varela}}, \ and\ \bibinfo {author} {\bibfnamefont {J.}~\bibnamefont
  {Santamaria}},\ }\href {\doibase 10.1038/nnano.2017.51} {\bibfield  {journal}
  {\bibinfo  {journal} {Nature Nanotechnology}\ }\textbf {\bibinfo {volume}
  {12}},\ \bibinfo {pages} {655–662} (\bibinfo {year} {2017})}\BibitemShut
  {NoStop}%
\bibitem [{\citenamefont {Blakesley}\ \emph {et~al.}(2005)\citenamefont
  {Blakesley}, \citenamefont {See}, \citenamefont {Shields}, \citenamefont
  {Kardynał}, \citenamefont {Atkinson}, \citenamefont {Farrer},\ and\
  \citenamefont {Ritchie}}]{Blakesley2005}%
  \BibitemOpen
  \bibfield  {author} {\bibinfo {author} {\bibfnamefont {J.~C.}\ \bibnamefont
  {Blakesley}}, \bibinfo {author} {\bibfnamefont {P.}~\bibnamefont {See}},
  \bibinfo {author} {\bibfnamefont {A.~J.}\ \bibnamefont {Shields}}, \bibinfo
  {author} {\bibfnamefont {B.~E.}\ \bibnamefont {Kardynał}}, \bibinfo {author}
  {\bibfnamefont {P.}~\bibnamefont {Atkinson}}, \bibinfo {author}
  {\bibfnamefont {I.}~\bibnamefont {Farrer}}, \ and\ \bibinfo {author}
  {\bibfnamefont {D.~A.}\ \bibnamefont {Ritchie}},\ }\href {\doibase
  10.1103/PhysRevLett.94.067401} {\bibfield  {journal} {\bibinfo  {journal}
  {Physical Review Letters}\ }\textbf {\bibinfo {volume} {94}},\ \bibinfo
  {pages} {067401} (\bibinfo {year} {2005})}\BibitemShut {NoStop}%
\bibitem [{\citenamefont {Faist}\ \emph {et~al.}(1994)\citenamefont {Faist},
  \citenamefont {Capasso}, \citenamefont {Sivco}, \citenamefont {Sirtori},
  \citenamefont {Hutchinson},\ and\ \citenamefont {Cho}}]{Faist1994}%
  \BibitemOpen
  \bibfield  {author} {\bibinfo {author} {\bibfnamefont {J.}~\bibnamefont
  {Faist}}, \bibinfo {author} {\bibfnamefont {F.}~\bibnamefont {Capasso}},
  \bibinfo {author} {\bibfnamefont {D.~L.}\ \bibnamefont {Sivco}}, \bibinfo
  {author} {\bibfnamefont {C.}~\bibnamefont {Sirtori}}, \bibinfo {author}
  {\bibfnamefont {A.~L.}\ \bibnamefont {Hutchinson}}, \ and\ \bibinfo {author}
  {\bibfnamefont {A.~Y.}\ \bibnamefont {Cho}},\ }\href {\doibase
  10.1126/science.264.5158.553} {\bibfield  {journal} {\bibinfo  {journal}
  {Science}\ }\textbf {\bibinfo {volume} {264}},\ \bibinfo {pages} {553–556}
  (\bibinfo {year} {1994})}\BibitemShut {NoStop}%
\bibitem [{\citenamefont {Yu}\ \emph {et~al.}(2021)\citenamefont {Yu},
  \citenamefont {Suzuki}, \citenamefont {Van~Ta}, \citenamefont {Suzuki},\ and\
  \citenamefont {Asada}}]{Yu2021}%
  \BibitemOpen
  \bibfield  {author} {\bibinfo {author} {\bibfnamefont {X.}~\bibnamefont
  {Yu}}, \bibinfo {author} {\bibfnamefont {Y.}~\bibnamefont {Suzuki}}, \bibinfo
  {author} {\bibfnamefont {M.}~\bibnamefont {Van~Ta}}, \bibinfo {author}
  {\bibfnamefont {S.}~\bibnamefont {Suzuki}}, \ and\ \bibinfo {author}
  {\bibfnamefont {M.}~\bibnamefont {Asada}},\ }\href {\doibase
  10.1109/LED.2021.3082577} {\bibfield  {journal} {\bibinfo  {journal} {IEEE
  Electron Device Letters}\ }\textbf {\bibinfo {volume} {42}},\ \bibinfo
  {pages} {982–985} (\bibinfo {year} {2021})}\BibitemShut {NoStop}%
\bibitem [{\citenamefont {Asada}\ and\ \citenamefont
  {Suzuki}(2016)}]{Asada2016}%
  \BibitemOpen
  \bibfield  {author} {\bibinfo {author} {\bibfnamefont {M.}~\bibnamefont
  {Asada}}\ and\ \bibinfo {author} {\bibfnamefont {S.}~\bibnamefont {Suzuki}},\
  }\href {\doibase 10.1007/s10762-016-0321-6} {\bibfield  {journal} {\bibinfo
  {journal} {Journal of Infrared, Millimeter, and Terahertz Waves}\ }\textbf
  {\bibinfo {volume} {37}},\ \bibinfo {pages} {1185} (\bibinfo {year}
  {2016})}\BibitemShut {NoStop}%
\bibitem [{\citenamefont {Xing}\ \emph {et~al.}(2019)\citenamefont {Xing},
  \citenamefont {Encomendero},\ and\ \citenamefont {Jena}}]{Xing2019}%
  \BibitemOpen
  \bibfield  {author} {\bibinfo {author} {\bibfnamefont {H.~G.}\ \bibnamefont
  {Xing}}, \bibinfo {author} {\bibfnamefont {J.}~\bibnamefont {Encomendero}}, \
  and\ \bibinfo {author} {\bibfnamefont {D.}~\bibnamefont {Jena}},\ }in\ \href
  {\doibase 10.1117/12.2512638} {\emph {\bibinfo {booktitle} {Gallium Nitride
  Materials and Devices XIV}}},\ Vol.\ \bibinfo {volume} {10918},\ \bibinfo
  {organization} {International Society for Optics and Photonics}\ (\bibinfo
  {publisher} {SPIE},\ \bibinfo {year} {2019})\ pp.\ \bibinfo {pages} {45 --
  50},\ \bibinfo {note} {doi:
  \href{https://doi.org/10.1117/12.2512638}{10.1117/12.2512638}}\BibitemShut
  {NoStop}%
\bibitem [{\citenamefont {Cho}\ \emph {et~al.}(2020)\citenamefont {Cho},
  \citenamefont {Encomendero}, \citenamefont {Ho}, \citenamefont {Xing},\ and\
  \citenamefont {Jena}}]{Cho2020}%
  \BibitemOpen
  \bibfield  {author} {\bibinfo {author} {\bibfnamefont {Y.}~\bibnamefont
  {Cho}}, \bibinfo {author} {\bibfnamefont {J.}~\bibnamefont {Encomendero}},
  \bibinfo {author} {\bibfnamefont {S.-T.}\ \bibnamefont {Ho}}, \bibinfo
  {author} {\bibfnamefont {H.~G.}\ \bibnamefont {Xing}}, \ and\ \bibinfo
  {author} {\bibfnamefont {D.}~\bibnamefont {Jena}},\ }\href {\doibase
  10.1063/5.0022143} {\bibfield  {journal} {\bibinfo  {journal} {Applied
  Physics Letters}\ }\textbf {\bibinfo {volume} {117}},\ \bibinfo {pages}
  {143501} (\bibinfo {year} {2020})}\BibitemShut {NoStop}%
\bibitem [{\citenamefont {Bonnefoi}\ \emph {et~al.}(1985)\citenamefont
  {Bonnefoi}, \citenamefont {McGill},\ and\ \citenamefont
  {Burnham}}]{Bonnefoi1985}%
  \BibitemOpen
  \bibfield  {author} {\bibinfo {author} {\bibfnamefont {A.}~\bibnamefont
  {Bonnefoi}}, \bibinfo {author} {\bibfnamefont {T.}~\bibnamefont {McGill}}, \
  and\ \bibinfo {author} {\bibfnamefont {R.}~\bibnamefont {Burnham}},\ }\href
  {\doibase 10.1109/EDL.1985.26258} {\bibfield  {journal} {\bibinfo  {journal}
  {IEEE Electron Device Letters}\ }\textbf {\bibinfo {volume} {6}},\ \bibinfo
  {pages} {636–638} (\bibinfo {year} {1985})}\BibitemShut {NoStop}%
\bibitem [{\citenamefont {Lind}\ \emph {et~al.}(2004)\citenamefont {Lind},
  \citenamefont {Lindstrom},\ and\ \citenamefont {Wernersson}}]{Lind2004}%
  \BibitemOpen
  \bibfield  {author} {\bibinfo {author} {\bibfnamefont {E.}~\bibnamefont
  {Lind}}, \bibinfo {author} {\bibfnamefont {P.}~\bibnamefont {Lindstrom}}, \
  and\ \bibinfo {author} {\bibfnamefont {L.-E.}\ \bibnamefont {Wernersson}},\
  }\href {\doibase 10.1109/LED.2004.835159} {\bibfield  {journal} {\bibinfo
  {journal} {IEEE Electron Device Letters}\ }\textbf {\bibinfo {volume} {25}},\
  \bibinfo {pages} {678–680} (\bibinfo {year} {2004})}\BibitemShut {NoStop}%
\bibitem [{\citenamefont {Condori~Quispe}\ \emph {et~al.}(2016)\citenamefont
  {Condori~Quispe}, \citenamefont {Encomendero-Risco}, \citenamefont {Xing},\
  and\ \citenamefont {Sensale-Rodriguez}}]{Condori2016}%
  \BibitemOpen
  \bibfield  {author} {\bibinfo {author} {\bibfnamefont {H.~O.}\ \bibnamefont
  {Condori~Quispe}}, \bibinfo {author} {\bibfnamefont {J.~J.}\ \bibnamefont
  {Encomendero-Risco}}, \bibinfo {author} {\bibfnamefont {H.~G.}\ \bibnamefont
  {Xing}}, \ and\ \bibinfo {author} {\bibfnamefont {B.}~\bibnamefont
  {Sensale-Rodriguez}},\ }\href {\doibase 10.1063/1.4961053} {\bibfield
  {journal} {\bibinfo  {journal} {Applied Physics Letters}\ }\textbf {\bibinfo
  {volume} {109}},\ \bibinfo {pages} {063111} (\bibinfo {year}
  {2016})}\BibitemShut {NoStop}%
\bibitem [{\citenamefont {King-Smith}\ and\ \citenamefont
  {Vanderbilt}(1993)}]{King1993}%
  \BibitemOpen
  \bibfield  {author} {\bibinfo {author} {\bibfnamefont {R.~D.}\ \bibnamefont
  {King-Smith}}\ and\ \bibinfo {author} {\bibfnamefont {D.}~\bibnamefont
  {Vanderbilt}},\ }\href {\doibase 10.1103/PhysRevB.47.1651} {\bibfield
  {journal} {\bibinfo  {journal} {Phys. Rev. B}\ }\textbf {\bibinfo {volume}
  {47}},\ \bibinfo {pages} {1651–1654} (\bibinfo {year} {1993})}\BibitemShut
  {NoStop}%
\bibitem [{\citenamefont {Resta}(1992)}]{Resta1992}%
  \BibitemOpen
  \bibfield  {author} {\bibinfo {author} {\bibfnamefont {R.}~\bibnamefont
  {Resta}},\ }\href {\doibase 10.1080/00150199208016065} {\bibfield  {journal}
  {\bibinfo  {journal} {Ferroelectrics}\ }\textbf {\bibinfo {volume} {136}},\
  \bibinfo {pages} {51–55} (\bibinfo {year} {1992})}\BibitemShut {NoStop}%
\bibitem [{\citenamefont {Encomendero}\ \emph {et~al.}(2019)\citenamefont
  {Encomendero}, \citenamefont {Protasenko}, \citenamefont {Sensale-Rodriguez},
  \citenamefont {Fay}, \citenamefont {Rana}, \citenamefont {Jena},\ and\
  \citenamefont {Xing}}]{Encomendero2019}%
  \BibitemOpen
  \bibfield  {author} {\bibinfo {author} {\bibfnamefont {J.}~\bibnamefont
  {Encomendero}}, \bibinfo {author} {\bibfnamefont {V.}~\bibnamefont
  {Protasenko}}, \bibinfo {author} {\bibfnamefont {B.}~\bibnamefont
  {Sensale-Rodriguez}}, \bibinfo {author} {\bibfnamefont {P.}~\bibnamefont
  {Fay}}, \bibinfo {author} {\bibfnamefont {F.}~\bibnamefont {Rana}}, \bibinfo
  {author} {\bibfnamefont {D.}~\bibnamefont {Jena}}, \ and\ \bibinfo {author}
  {\bibfnamefont {H.~G.}\ \bibnamefont {Xing}},\ }\href {\doibase
  10.1103/PhysRevApplied.11.034032} {\bibfield  {journal} {\bibinfo  {journal}
  {Phys. Rev. Applied}\ }\textbf {\bibinfo {volume} {11}},\ \bibinfo {pages}
  {034032} (\bibinfo {year} {2019})}\BibitemShut {NoStop}%
\bibitem [{\citenamefont {Ricco}\ and\ \citenamefont
  {Azbel}(1984)}]{Ricco1984}%
  \BibitemOpen
  \bibfield  {author} {\bibinfo {author} {\bibfnamefont {B.}~\bibnamefont
  {Ricco}}\ and\ \bibinfo {author} {\bibfnamefont {M.~Y.}\ \bibnamefont
  {Azbel}},\ }\href {\doibase 10.1103/PhysRevB.29.1970} {\bibfield  {journal}
  {\bibinfo  {journal} {Physical Review B}\ }\textbf {\bibinfo {volume} {29}},\
  \bibinfo {pages} {1970–1981} (\bibinfo {year} {1984})}\BibitemShut
  {NoStop}%
\bibitem [{\citenamefont {Buttiker}(1988)}]{Buttiker1988}%
  \BibitemOpen
  \bibfield  {author} {\bibinfo {author} {\bibfnamefont {M.}~\bibnamefont
  {Buttiker}},\ }\href@noop {} {\bibfield  {journal} {\bibinfo  {journal} {IBM
  Journal of Research and Development}\ }\textbf {\bibinfo {volume} {32}},\
  \bibinfo {pages} {63} (\bibinfo {year} {1988})}\BibitemShut {NoStop}%
\bibitem [{\citenamefont {Schubert}(1994)}]{Schubert1994_1}%
  \BibitemOpen
  \bibfield  {author} {\bibinfo {author} {\bibfnamefont {E.}~\bibnamefont
  {Schubert}},\ }in\ \href {\doibase
  https://doi.org/10.1016/S0080-8784(08)62662-9} {\emph {\bibinfo {booktitle}
  {Epitaxial Microstructures}}},\ \bibinfo {series} {Semiconductors and
  Semimetals}, Vol.~\bibinfo {volume} {40},\ \bibinfo {editor} {edited by\
  \bibinfo {editor} {\bibfnamefont {A.~C.}\ \bibnamefont {Gossard}}}\ (\bibinfo
   {publisher} {Elsevier},\ \bibinfo {year} {1994})\ pp.\ \bibinfo {pages}
  {1--151}\BibitemShut {NoStop}%
\bibitem [{\citenamefont {Encomendero}\ \emph {et~al.}(2018)\citenamefont
  {Encomendero}, \citenamefont {Yan}, \citenamefont {Verma}, \citenamefont
  {Islam}, \citenamefont {Protasenko}, \citenamefont {Rouvimov}, \citenamefont
  {Fay}, \citenamefont {Jena},\ and\ \citenamefont {Xing}}]{Encomendero2018}%
  \BibitemOpen
  \bibfield  {author} {\bibinfo {author} {\bibfnamefont {J.}~\bibnamefont
  {Encomendero}}, \bibinfo {author} {\bibfnamefont {R.}~\bibnamefont {Yan}},
  \bibinfo {author} {\bibfnamefont {A.}~\bibnamefont {Verma}}, \bibinfo
  {author} {\bibfnamefont {S.~M.}\ \bibnamefont {Islam}}, \bibinfo {author}
  {\bibfnamefont {V.}~\bibnamefont {Protasenko}}, \bibinfo {author}
  {\bibfnamefont {S.}~\bibnamefont {Rouvimov}}, \bibinfo {author}
  {\bibfnamefont {P.}~\bibnamefont {Fay}}, \bibinfo {author} {\bibfnamefont
  {D.}~\bibnamefont {Jena}}, \ and\ \bibinfo {author} {\bibfnamefont {H.~G.}\
  \bibnamefont {Xing}},\ }\href {\doibase 10.1063/1.5016414} {\bibfield
  {journal} {\bibinfo  {journal} {Applied Physics Letters}\ }\textbf {\bibinfo
  {volume} {112}},\ \bibinfo {pages} {103101} (\bibinfo {year}
  {2018})}\BibitemShut {NoStop}%
\bibitem [{\citenamefont {Encomendero}\ \emph
  {et~al.}(2020{\natexlab{a}})\citenamefont {Encomendero}, \citenamefont
  {Jena},\ and\ \citenamefont {Xing}}]{Encomendero2020}%
  \BibitemOpen
  \bibfield  {author} {\bibinfo {author} {\bibfnamefont {J.}~\bibnamefont
  {Encomendero}}, \bibinfo {author} {\bibfnamefont {D.}~\bibnamefont {Jena}}, \
  and\ \bibinfo {author} {\bibfnamefont {H.~G.}\ \bibnamefont {Xing}},\
  }\enquote {\bibinfo {title} {{Resonant Tunneling Transport in Polar
  III-Nitride Heterostructures}},}\ in\ \href {\doibase
  10.1007/978-3-030-20208-8_8} {\emph {\bibinfo {booktitle} {High-Frequency GaN
  Electronic Devices}}}\ (\bibinfo  {publisher} {Springer International
  Publishing},\ \bibinfo {address} {Cham},\ \bibinfo {year} {2020})\ pp.\
  \bibinfo {pages} {215--247}\BibitemShut {NoStop}%
\bibitem [{\citenamefont {Encomendero}\ \emph
  {et~al.}(2021{\natexlab{a}})\citenamefont {Encomendero}, \citenamefont
  {Islam}, \citenamefont {Jena},\ and\ \citenamefont
  {Grace~Xing}}]{Encomendero2021_JVSTA}%
  \BibitemOpen
  \bibfield  {author} {\bibinfo {author} {\bibfnamefont {J.}~\bibnamefont
  {Encomendero}}, \bibinfo {author} {\bibfnamefont {S.}~\bibnamefont {Islam}},
  \bibinfo {author} {\bibfnamefont {D.}~\bibnamefont {Jena}}, \ and\ \bibinfo
  {author} {\bibfnamefont {H.}~\bibnamefont {Grace~Xing}},\ }\href {\doibase
  10.1116/6.0000775} {\bibfield  {journal} {\bibinfo  {journal} {Journal of
  Vacuum Science \& Technology A}\ }\textbf {\bibinfo {volume} {39}},\ \bibinfo
  {pages} {023409} (\bibinfo {year} {2021}{\natexlab{a}})}\BibitemShut
  {NoStop}%
\bibitem [{\citenamefont {Encomendero}\ \emph {et~al.}(2017)\citenamefont
  {Encomendero}, \citenamefont {Faria}, \citenamefont {Islam}, \citenamefont
  {Protasenko}, \citenamefont {Rouvimov}, \citenamefont {Sensale-Rodriguez},
  \citenamefont {Fay}, \citenamefont {Jena},\ and\ \citenamefont
  {Xing}}]{Encomendero2017}%
  \BibitemOpen
  \bibfield  {author} {\bibinfo {author} {\bibfnamefont {J.}~\bibnamefont
  {Encomendero}}, \bibinfo {author} {\bibfnamefont {F.~A.}\ \bibnamefont
  {Faria}}, \bibinfo {author} {\bibfnamefont {S.~M.}\ \bibnamefont {Islam}},
  \bibinfo {author} {\bibfnamefont {V.}~\bibnamefont {Protasenko}}, \bibinfo
  {author} {\bibfnamefont {S.}~\bibnamefont {Rouvimov}}, \bibinfo {author}
  {\bibfnamefont {B.}~\bibnamefont {Sensale-Rodriguez}}, \bibinfo {author}
  {\bibfnamefont {P.}~\bibnamefont {Fay}}, \bibinfo {author} {\bibfnamefont
  {D.}~\bibnamefont {Jena}}, \ and\ \bibinfo {author} {\bibfnamefont {H.~G.}\
  \bibnamefont {Xing}},\ }\href {\doibase 10.1103/PhysRevX.7.041017} {\bibfield
   {journal} {\bibinfo  {journal} {Phys. Rev. X}\ }\textbf {\bibinfo {volume}
  {7}},\ \bibinfo {pages} {041017} (\bibinfo {year} {2017})}\BibitemShut
  {NoStop}%
\bibitem [{\citenamefont {Encomendero}\ \emph
  {et~al.}(2021{\natexlab{b}})\citenamefont {Encomendero}, \citenamefont
  {Protasenko}, \citenamefont {Jena},\ and\ \citenamefont
  {Xing}}]{Encomendero2021APEX}%
  \BibitemOpen
  \bibfield  {author} {\bibinfo {author} {\bibfnamefont {J.}~\bibnamefont
  {Encomendero}}, \bibinfo {author} {\bibfnamefont {V.}~\bibnamefont
  {Protasenko}}, \bibinfo {author} {\bibfnamefont {D.}~\bibnamefont {Jena}}, \
  and\ \bibinfo {author} {\bibfnamefont {H.~G.}\ \bibnamefont {Xing}},\ }\href
  {\doibase 10.35848/1882-0786/ac345e} {\bibfield  {journal} {\bibinfo
  {journal} {Applied Physics Express}\ }\textbf {\bibinfo {volume} {14}},\
  \bibinfo {pages} {122003} (\bibinfo {year} {2021}{\natexlab{b}})}\BibitemShut
  {NoStop}%
\bibitem [{\citenamefont {{Encomendero}}\ \emph {et~al.}(2022)\citenamefont
  {{Encomendero}}, \citenamefont {{Protasenko}}, \citenamefont {{Jena}},\ and\
  \citenamefont {{Xing}}}]{Encomendero2022APS}%
  \BibitemOpen
  \bibfield  {author} {\bibinfo {author} {\bibfnamefont {J.}~\bibnamefont
  {{Encomendero}}}, \bibinfo {author} {\bibfnamefont {V.}~\bibnamefont
  {{Protasenko}}}, \bibinfo {author} {\bibfnamefont {D.}~\bibnamefont
  {{Jena}}}, \ and\ \bibinfo {author} {\bibfnamefont {G.}~\bibnamefont
  {{Xing}}},\ }in\ \href@noop {} {\emph {\bibinfo {booktitle} {APS March
  Meeting Abstracts}}},\ \bibinfo {series} {APS Meeting Abstracts}, Vol.\
  \bibinfo {volume} {2022}\ (\bibinfo {year} {2022})\ p.\ \bibinfo {pages}
  {Q68.005}\BibitemShut {NoStop}%
\bibitem [{\citenamefont {Growden}\ \emph {et~al.}(2020)\citenamefont
  {Growden}, \citenamefont {Storm}, \citenamefont {Cornuelle}, \citenamefont
  {Brown}, \citenamefont {Zhang}, \citenamefont {Downey}, \citenamefont
  {Roussos}, \citenamefont {Cronk}, \citenamefont {Ruppalt}, \citenamefont
  {Champlain}, \citenamefont {Berger},\ and\ \citenamefont
  {Meyer}}]{Growden2020}%
  \BibitemOpen
  \bibfield  {author} {\bibinfo {author} {\bibfnamefont {T.~A.}\ \bibnamefont
  {Growden}}, \bibinfo {author} {\bibfnamefont {D.~F.}\ \bibnamefont {Storm}},
  \bibinfo {author} {\bibfnamefont {E.~M.}\ \bibnamefont {Cornuelle}}, \bibinfo
  {author} {\bibfnamefont {E.~R.}\ \bibnamefont {Brown}}, \bibinfo {author}
  {\bibfnamefont {W.}~\bibnamefont {Zhang}}, \bibinfo {author} {\bibfnamefont
  {B.~P.}\ \bibnamefont {Downey}}, \bibinfo {author} {\bibfnamefont {J.~A.}\
  \bibnamefont {Roussos}}, \bibinfo {author} {\bibfnamefont {N.}~\bibnamefont
  {Cronk}}, \bibinfo {author} {\bibfnamefont {L.~B.}\ \bibnamefont {Ruppalt}},
  \bibinfo {author} {\bibfnamefont {J.~G.}\ \bibnamefont {Champlain}}, \bibinfo
  {author} {\bibfnamefont {P.~R.}\ \bibnamefont {Berger}}, \ and\ \bibinfo
  {author} {\bibfnamefont {D.~J.}\ \bibnamefont {Meyer}},\ }\href {\doibase
  10.1063/1.5139219} {\bibfield  {journal} {\bibinfo  {journal} {Applied
  Physics Letters}\ }\textbf {\bibinfo {volume} {116}},\ \bibinfo {pages}
  {113501} (\bibinfo {year} {2020})}\BibitemShut {NoStop}%
\bibitem [{\citenamefont {Growden}\ \emph {et~al.}(2016)\citenamefont
  {Growden}, \citenamefont {Storm}, \citenamefont {Zhang}, \citenamefont
  {Brown}, \citenamefont {Meyer}, \citenamefont {Fakhimi},\ and\ \citenamefont
  {Berger}}]{Growden2016}%
  \BibitemOpen
  \bibfield  {author} {\bibinfo {author} {\bibfnamefont {T.~A.}\ \bibnamefont
  {Growden}}, \bibinfo {author} {\bibfnamefont {D.~F.}\ \bibnamefont {Storm}},
  \bibinfo {author} {\bibfnamefont {W.}~\bibnamefont {Zhang}}, \bibinfo
  {author} {\bibfnamefont {E.~R.}\ \bibnamefont {Brown}}, \bibinfo {author}
  {\bibfnamefont {D.~J.}\ \bibnamefont {Meyer}}, \bibinfo {author}
  {\bibfnamefont {P.}~\bibnamefont {Fakhimi}}, \ and\ \bibinfo {author}
  {\bibfnamefont {P.~R.}\ \bibnamefont {Berger}},\ }\href
  {http://scitation.aip.org/content/aip/journal/apl/109/8/10.1063/1.4961442}
  {\bibfield  {journal} {\bibinfo  {journal} {Applied Physics Letters}\
  }\textbf {\bibinfo {volume} {109}},\ \bibinfo {eid} {083504} (\bibinfo {year}
  {2016})}\BibitemShut {NoStop}%
\bibitem [{\citenamefont {Encomendero}\ \emph
  {et~al.}(2020{\natexlab{b}})\citenamefont {Encomendero}, \citenamefont
  {Protasenko}, \citenamefont {Rana}, \citenamefont {Jena},\ and\ \citenamefont
  {Xing}}]{Encomendero2020PRA}%
  \BibitemOpen
  \bibfield  {author} {\bibinfo {author} {\bibfnamefont {J.}~\bibnamefont
  {Encomendero}}, \bibinfo {author} {\bibfnamefont {V.}~\bibnamefont
  {Protasenko}}, \bibinfo {author} {\bibfnamefont {F.}~\bibnamefont {Rana}},
  \bibinfo {author} {\bibfnamefont {D.}~\bibnamefont {Jena}}, \ and\ \bibinfo
  {author} {\bibfnamefont {H.~G.}\ \bibnamefont {Xing}},\ }\href {\doibase
  10.1103/PhysRevApplied.13.034048} {\bibfield  {journal} {\bibinfo  {journal}
  {Phys. Rev. Applied}\ }\textbf {\bibinfo {volume} {13}},\ \bibinfo {pages}
  {034048} (\bibinfo {year} {2020}{\natexlab{b}})}\BibitemShut {NoStop}%
\bibitem [{\citenamefont {Growden}\ \emph {et~al.}(2018)\citenamefont
  {Growden}, \citenamefont {Zhang}, \citenamefont {Brown}, \citenamefont
  {Storm}, \citenamefont {Hansen}, \citenamefont {Fakhimi}, \citenamefont
  {Meyer},\ and\ \citenamefont {Berger}}]{Growden2018}%
  \BibitemOpen
  \bibfield  {author} {\bibinfo {author} {\bibfnamefont {T.~A.}\ \bibnamefont
  {Growden}}, \bibinfo {author} {\bibfnamefont {W.}~\bibnamefont {Zhang}},
  \bibinfo {author} {\bibfnamefont {E.~R.}\ \bibnamefont {Brown}}, \bibinfo
  {author} {\bibfnamefont {D.~F.}\ \bibnamefont {Storm}}, \bibinfo {author}
  {\bibfnamefont {K.}~\bibnamefont {Hansen}}, \bibinfo {author} {\bibfnamefont
  {P.}~\bibnamefont {Fakhimi}}, \bibinfo {author} {\bibfnamefont {D.~J.}\
  \bibnamefont {Meyer}}, \ and\ \bibinfo {author} {\bibfnamefont {P.~R.}\
  \bibnamefont {Berger}},\ }\href {\doibase 10.1063/1.5010794} {\bibfield
  {journal} {\bibinfo  {journal} {Applied Physics Letters}\ }\textbf {\bibinfo
  {volume} {112}},\ \bibinfo {pages} {033508} (\bibinfo {year}
  {2018})}\BibitemShut {NoStop}%
\bibitem [{\citenamefont {Growden}\ \emph {et~al.}(2019)\citenamefont
  {Growden}, \citenamefont {Cornuelle}, \citenamefont {Storm}, \citenamefont
  {Zhang}, \citenamefont {Brown}, \citenamefont {Whitaker}, \citenamefont
  {Daulton}, \citenamefont {Molnar}, \citenamefont {Meyer},\ and\ \citenamefont
  {Berger}}]{Growden2019}%
  \BibitemOpen
  \bibfield  {author} {\bibinfo {author} {\bibfnamefont {T.~A.}\ \bibnamefont
  {Growden}}, \bibinfo {author} {\bibfnamefont {E.~M.}\ \bibnamefont
  {Cornuelle}}, \bibinfo {author} {\bibfnamefont {D.~F.}\ \bibnamefont
  {Storm}}, \bibinfo {author} {\bibfnamefont {W.}~\bibnamefont {Zhang}},
  \bibinfo {author} {\bibfnamefont {E.~R.}\ \bibnamefont {Brown}}, \bibinfo
  {author} {\bibfnamefont {L.~M.}\ \bibnamefont {Whitaker}}, \bibinfo {author}
  {\bibfnamefont {J.~W.}\ \bibnamefont {Daulton}}, \bibinfo {author}
  {\bibfnamefont {R.}~\bibnamefont {Molnar}}, \bibinfo {author} {\bibfnamefont
  {D.~J.}\ \bibnamefont {Meyer}}, \ and\ \bibinfo {author} {\bibfnamefont
  {P.~R.}\ \bibnamefont {Berger}},\ }\href {\doibase 10.1063/1.5095056}
  {\bibfield  {journal} {\bibinfo  {journal} {Applied Physics Letters}\
  }\textbf {\bibinfo {volume} {114}},\ \bibinfo {pages} {203503} (\bibinfo
  {year} {2019})}\BibitemShut {NoStop}%
\bibitem [{\citenamefont {Zhang}\ \emph
  {et~al.}(2021{\natexlab{a}})\citenamefont {Zhang}, \citenamefont {Xue},
  \citenamefont {Fu}, \citenamefont {Li}, \citenamefont {Sun}, \citenamefont
  {Yao}, \citenamefont {Liu}, \citenamefont {Zhang}, \citenamefont {Ma},
  \citenamefont {Zhang},\ and\ \citenamefont {Hao}}]{Zhang2021JAP}%
  \BibitemOpen
  \bibfield  {author} {\bibinfo {author} {\bibfnamefont {H.}~\bibnamefont
  {Zhang}}, \bibinfo {author} {\bibfnamefont {J.}~\bibnamefont {Xue}}, \bibinfo
  {author} {\bibfnamefont {Y.}~\bibnamefont {Fu}}, \bibinfo {author}
  {\bibfnamefont {L.}~\bibnamefont {Li}}, \bibinfo {author} {\bibfnamefont
  {Z.}~\bibnamefont {Sun}}, \bibinfo {author} {\bibfnamefont {J.}~\bibnamefont
  {Yao}}, \bibinfo {author} {\bibfnamefont {F.}~\bibnamefont {Liu}}, \bibinfo
  {author} {\bibfnamefont {K.}~\bibnamefont {Zhang}}, \bibinfo {author}
  {\bibfnamefont {X.}~\bibnamefont {Ma}}, \bibinfo {author} {\bibfnamefont
  {J.}~\bibnamefont {Zhang}}, \ and\ \bibinfo {author} {\bibfnamefont
  {Y.}~\bibnamefont {Hao}},\ }\href {\doibase 10.1063/5.0033324} {\bibfield
  {journal} {\bibinfo  {journal} {Journal of Applied Physics}\ }\textbf
  {\bibinfo {volume} {129}},\ \bibinfo {pages} {014502} (\bibinfo {year}
  {2021}{\natexlab{a}})}\BibitemShut {NoStop}%
\bibitem [{\citenamefont {Zhang}\ \emph
  {et~al.}(2021{\natexlab{b}})\citenamefont {Zhang}, \citenamefont {Xue},
  \citenamefont {Sun}, \citenamefont {Li}, \citenamefont {Yao}, \citenamefont
  {Liu}, \citenamefont {Yang}, \citenamefont {Wu}, \citenamefont {Li},
  \citenamefont {Fu}, \citenamefont {Liu}, \citenamefont {Zhang},\ and\
  \citenamefont {Hao}}]{Zhang2021APL}%
  \BibitemOpen
  \bibfield  {author} {\bibinfo {author} {\bibfnamefont {H.}~\bibnamefont
  {Zhang}}, \bibinfo {author} {\bibfnamefont {J.}~\bibnamefont {Xue}}, \bibinfo
  {author} {\bibfnamefont {Z.}~\bibnamefont {Sun}}, \bibinfo {author}
  {\bibfnamefont {L.}~\bibnamefont {Li}}, \bibinfo {author} {\bibfnamefont
  {J.}~\bibnamefont {Yao}}, \bibinfo {author} {\bibfnamefont {F.}~\bibnamefont
  {Liu}}, \bibinfo {author} {\bibfnamefont {X.}~\bibnamefont {Yang}}, \bibinfo
  {author} {\bibfnamefont {G.}~\bibnamefont {Wu}}, \bibinfo {author}
  {\bibfnamefont {Z.}~\bibnamefont {Li}}, \bibinfo {author} {\bibfnamefont
  {Y.}~\bibnamefont {Fu}}, \bibinfo {author} {\bibfnamefont {Z.}~\bibnamefont
  {Liu}}, \bibinfo {author} {\bibfnamefont {J.}~\bibnamefont {Zhang}}, \ and\
  \bibinfo {author} {\bibfnamefont {Y.}~\bibnamefont {Hao}},\ }\href {\doibase
  10.1063/5.0064790} {\bibfield  {journal} {\bibinfo  {journal} {Applied
  Physics Letters}\ }\textbf {\bibinfo {volume} {119}},\ \bibinfo {pages}
  {153506} (\bibinfo {year} {2021}{\natexlab{b}})}\BibitemShut {NoStop}%
\bibitem [{\citenamefont {Wang}\ \emph {et~al.}(2018)\citenamefont {Wang},
  \citenamefont {Su}, \citenamefont {Chen}, \citenamefont {Wang}, \citenamefont
  {Yang}, \citenamefont {Sheng}, \citenamefont {Lin}, \citenamefont {Rong},
  \citenamefont {Wang}, \citenamefont {Shi}, \citenamefont {Tan}, \citenamefont
  {Zhang}, \citenamefont {Ge}, \citenamefont {Shen}, \citenamefont {Liu},\ and\
  \citenamefont {Wang}}]{Wang2018}%
  \BibitemOpen
  \bibfield  {author} {\bibinfo {author} {\bibfnamefont {D.}~\bibnamefont
  {Wang}}, \bibinfo {author} {\bibfnamefont {J.}~\bibnamefont {Su}}, \bibinfo
  {author} {\bibfnamefont {Z.}~\bibnamefont {Chen}}, \bibinfo {author}
  {\bibfnamefont {T.}~\bibnamefont {Wang}}, \bibinfo {author} {\bibfnamefont
  {L.}~\bibnamefont {Yang}}, \bibinfo {author} {\bibfnamefont {B.}~\bibnamefont
  {Sheng}}, \bibinfo {author} {\bibfnamefont {S.}~\bibnamefont {Lin}}, \bibinfo
  {author} {\bibfnamefont {X.}~\bibnamefont {Rong}}, \bibinfo {author}
  {\bibfnamefont {P.}~\bibnamefont {Wang}}, \bibinfo {author} {\bibfnamefont
  {X.}~\bibnamefont {Shi}}, \bibinfo {author} {\bibfnamefont {W.}~\bibnamefont
  {Tan}}, \bibinfo {author} {\bibfnamefont {J.}~\bibnamefont {Zhang}}, \bibinfo
  {author} {\bibfnamefont {W.}~\bibnamefont {Ge}}, \bibinfo {author}
  {\bibfnamefont {B.}~\bibnamefont {Shen}}, \bibinfo {author} {\bibfnamefont
  {Y.}~\bibnamefont {Liu}}, \ and\ \bibinfo {author} {\bibfnamefont
  {X.}~\bibnamefont {Wang}},\ }\href {\doibase 10.1002/aelm.201800651}
  {\bibfield  {journal} {\bibinfo  {journal} {Advanced Electronic Materials}\
  }\textbf {\bibinfo {volume} {1800651}},\ \bibinfo {pages} {1800651} (\bibinfo
  {year} {2018})}\BibitemShut {NoStop}%
\bibitem [{\citenamefont {Song}\ \emph {et~al.}(2016)\citenamefont {Song},
  \citenamefont {Bhat}, \citenamefont {Bouzi}, \citenamefont {Zah},\ and\
  \citenamefont {Gmachl}}]{Song2016}%
  \BibitemOpen
  \bibfield  {author} {\bibinfo {author} {\bibfnamefont {A.~Y.}\ \bibnamefont
  {Song}}, \bibinfo {author} {\bibfnamefont {R.}~\bibnamefont {Bhat}}, \bibinfo
  {author} {\bibfnamefont {P.}~\bibnamefont {Bouzi}}, \bibinfo {author}
  {\bibfnamefont {C.-E.}\ \bibnamefont {Zah}}, \ and\ \bibinfo {author}
  {\bibfnamefont {C.~F.}\ \bibnamefont {Gmachl}},\ }\href {\doibase
  10.1103/PhysRevB.94.165307} {\bibfield  {journal} {\bibinfo  {journal}
  {Physical Review B}\ }\textbf {\bibinfo {volume} {94}},\ \bibinfo {pages}
  {165307} (\bibinfo {year} {2016})}\BibitemShut {NoStop}%
\end{thebibliography}%

\end{document}